% Choose 'big' when asked.%%%%%%%%%%%%%%%%%%
% To get double spaced version, search the word 'blaa' and comment out 
% the \baselineskip commands.

\input harvmac
\input epsf

%
%\def\listrefs{\footatend\immediate\closeout\rfile\writestoppt
%\baselineskip=11pt\centerline{{\bf References}}\bigskip{\frenchspacing%
%\parindent=20pt\escapechar=` \input refs.tmp\vfill\eject}\nonfrenchspacing}
%

%                        footnotes
\newskip\footskip\footskip8pt plus 1pt minus 1pt %sets footnote baselineskip
\def\footnotefont{\ninepoint}\def\f@t#1{\footnotefont #1\@foot}
\def\f@@t{\baselineskip\footskip\bgroup\footnotefont\aftergroup\@foot\let\next}
\setbox\strutbox=\hbox{\vrule height9.5pt depth4.5pt width0pt}
\global\newcount\ftno \global\ftno=0
\def\foot{\global\advance\ftno by1\footnote{$^{\the\ftno}$}}
%
%say \footend to put footnotes at end
%will cause problems if \ref used inside \foot, instead use \nref before
\newwrite\ftfile   
\def\footend{\def\foot{\global\advance\ftno by1\chardef\wfile=\ftfile
$^{\the\ftno}$\ifnum\ftno=1\immediate\openout\ftfile=foots.tmp\fi% 
\immediate\write\ftfile{\noexpand\smallskip%
\noexpand\item{f\the\ftno:\ }\pctsign}\findarg}%
\def\footatend{\vfill\eject\immediate\closeout\ftfile{\parindent=20pt   
\centerline{\bf Footnotes}\nobreak\bigskip\input foots.tmp }}}
\def\footatend{}

\baselineskip=10pt plus 1pt minus 1pt %blaa

\def\prl#1{{\it Phys. Rev. Lett.} {\bf #1}}
\def\prd#1{{\it Phys. Rev. D.} {\bf #1}}
\def\cmp#1{{\it Comm. Math. Phys.} {\bf #1}}
\def\mpl#1{{\it Mod. Phys. Lett. A.} {\bf #1}}
\def\plb#1{{\it Phys. Lett. B.} {\bf #1}}
\def\tilde{\widetilde}

\lref\wil{Shapere, A., Trivedi, S. and Wilczek, F. 1991, ``Dual Dilaton 
Dyons,'' \mpl{6}, 2677-2686.}

\lref\sz{Straumann, N. and Zhou, Z. H. 1990, ``Instability of the 
Bartnick-McKinnon Solution of the
Einstein-Yang-Mills Equations,'' \plb{237}, 353-356.}

\lref\heusler{Heusler, M. 1996, ``No-Hair Theorems and Black Holes with 
Hair'' (grqc 9610019)}

\lref\gidd{ Giddings, S.~B., Harvey, J.~A., Polchinski, J.~G., Shenker, 
S.~H., and Strominger, A. 1994, ``Hairy Black Holes in String Theory,''
\prd{50}, 6422-6426 (hepth 9309152).} 
\lref\wald{ Wald, R.~M. 1984, {\it General Relativity} (Chicago: 
University of   
Chicago Press), and references therein.} 
\lref\bek{  Bekenstein, J.~D. 1972, ``Transcendence 
of the Law of         
Baryon-Number Conservation in Black-Hole Physics,'' \prl{28}, 
452-455.     }
 \lref\adler{   Adler, S.~L., and Pearson, R.~B. 1978, 
``No-hair theorems for         
the Abelian Higgs and Goldstone models,'' \prd{18},   2798-2803.}
 \lref\bizon{  Bizo\'{n}, 
P. 1990, ``Colored Black Holes,'' \prl{64}, 2844-2847.  }
 \lref\smoller{ Smoller, J.~A., Wasserman, A.~G., and Yau, 
S.~T. 1993, 
``Existence of black hole solutions for the Einstein/Yang-Mills Equations,"
\cmp{154}, 377-401. }
 \lref\lee{
 Lee, K., Nair, V.~P., and Weinberg, E. 1992, ``Black Holes in         
Magnetic Monopoles,'' \prd{45}, 2751-2761. (hepth 9112008)  }
 \lref\kt{   Kastor, D., and Traschen, J. 1992, 
``Horizons Inside Classical         
Lumps,'' \prd{46}, 5399-5403. (hepth 9207070) }
\lref\bow{   Bowick, M.~J., Giddings, S.~
B., 
Harvey, J.~A., Horowitz, G.~T.,         
and Strominger, A. 1988, ``Axionic Black Holes and an Aharonov-Bohm Effect
for Strings,'' \prl{61}, 2823-2826.  }
 \lref\press{Press, W.~H., Teukolsky, S.~A., Vetterling, W.~T. and 
Flannery, B.~P  1992, {\it Numerical Recipes{\rm :} 
The        
Art of Scientific Computing} (Cambridge: Cambridge University Press).}
\lref\garfinkle{   Garfinkle,
D., Horowitz, G.~T., and Strominger, A.
1991, ``Charged
Black Holes in String Theory,'' \prd{43}, 3140-3143. erratum
 {\bf 45}, 3888, 1992.}

\Title{}{Instanton 
Supported Scalar Hair on Black Holes}

\centerline{Jennie Traschen\footnote{$^*$}{lboo@phast.umass.edu} and 
K.~Z.~Win\footnote{$^\dagger$}
{win@phast.umass.edu} }
\bigskip\centerline{\it Department of Physics and Astronomy}
\centerline{\it University of Massachusetts}\centerline{\it Amherst, MA 
01003}

\midinsert 
\narrower\narrower
\noindent

We 
present analytical perturbative,  and numerical solutions of the Einstein 
equation
which describe a black hole  with a nontrivial dilaton  field  
and a purely topological
gauge potential. The gauge 
potential has zero field strength
and hence no stress-energy, but it does couple to virtual string worldsheets
which wrap around the Euclidean horizon two-sphere, and generate an
effective interaction in the spacetime lagrangian. We use the 
 lagrangian   with
a nonstandard potential for a scalar field  that reproduces the effect of 
the worldsheet instantons. As has been previously pointed out the 
topological
charge $Q$ of the gauge field can be detected by an Anharonov-Bohm type 
experiment using
quantum strings; the classical scalar hair of the solutions here is a 
classical detection of $Q$.
ADM mass, dilaton charge and Hawking 
temperature are 
calculated and compared with the known cases when appropriate.
We discuss why these solutions do not violate the no-hair theorems and 
show that, for sufficiently small interaction coupling, the solutions are  
stable 
under linear time dependent perturbations.

\endinsert

\Date{February 1997}

\newsec{Introduction}

\noindent
The no-hair theorems in the physics of black holes essentially 
state that the
 only physical parameters describing a black hole are 
 its mass, angular momentum, and conserved charges associated with long 
range fields. 
Originally no-hair theorems arose from the uniqueness theorems of Israel, 
Carter
 and Wald
\wald \ for the Kerr solution.
If the cosmic censorship conjecture 
is correct, and hence the gravitational collapse of a star gives rise to 
a 
black hole, the no-hair theorems imply that the black hole  must be 
described 
by the Kerr solution.
An  alternative  method of proving a no-hair 
theorem, developed by Bekenstein \bek \  is
to show 
that the 
black hole geometry cannot support nontrivial classical field 
configurations that are of 
interest. We review this in section 5. In particular, a no-hair
theorem must be proved (or disproved) for each matter theory of interest.
For a scalar field coupled to gravity, there are simple proofs when 
either the
potential energy $V(\phi)$ or $\phi{\partial V/ \partial \phi}$ is 
positive definite 
 polynomial. So scalar hair is 
ruled
out for the standard self
interacting potential $V(\phi)=m^2\phi^2/2 +\lambda\phi^4/4$  
if $m^2,\lambda>0$.
The important case of spontaneous symmetry breaking, when $m^2<0$
and $\phi$ is coupled to a gauge field, requires a more
careful arguement, and is first proved in \adler.

In recent years, however, black holes with 
various kinds of hair have been discovered. The  ``colored'' black hole 
supporting Yang-Mills gauge field was  first numerically discovered  
\bizon, and existence 
was  later
proven analytically  \smoller. The colored black hole possesses zero 
Yang-Mills charge 
and thus parametrized only by its ADM mass.  Further,  it was shown 
\lee \ that a 
black hole can exist inside a magnetic monople of spontaneously broken gauge 
theories.  Under much more generality it was  shown  in \kt \ that black 
holes can exist inside various classical field 
configurations of which magnetic monopole is only a special case. 
The charged dilaton black holes \garfinkle \wil \ carry electric and 
magnetic charges,
and also have nontrivial scalar fields. Now, in all these examples of
black holes with scalar hair, the black hole also carries electric
and/or magnetic charges.
In this paper, we present numerical solutions,
and analytic perturbative solutions, for 
black holes with nontrivial dilaton field, but no electric or magnetic 
charge.
Instead, the black hole has a topological ``axion'' charge, that is, there
is a nontrivial 2-form gauge potential $B_{ab}$, with vanishing field
strength, that wraps around the horizon two-sphere. Classically $B_{ab}$
is not observable; its existence can be inferred by a distant observer,
by the nontrivial scalar field on the black hole.

These are asymptotically flat versions of the black holes constructed
by Giddings et.al.~\gidd\ and are the 
solutions to bosonic string theory, which correspond to black holes with 
a dilaton and
a purely topolgical axion gauge potential.
However, these exact solutions
are ``all throat,'' i.e.~the radius of two sphere is constant. The authors
write down a spacetime lagrangian which  contains an effective potential
giving the effect of Euclidean string world sheets that couple to the 
gauge potential and  wrap around
the horizon two-sphere. It is this 
interaction term which allows  one to get around the usual no hair 
construction. Hence the classically observable scalar hair is supported 
by the classically unobservable world sheet instantons.

In classical electrodynamics the gauge invariant field strength is the 
physically observable quantity.  However in 
quantum 
mechanics the potential can affect  a particle even in those regions where 
the 
field strength is identically zero, as illustrated in the 
famous Aharonov-Bohm effect.  A similar phenomenon comes up in string 
theory.   A string naturally couples to a 2-form potential, $B_{ab}$, 
since its trajectory is described by a world sheet.
   $H_{abc}(=\nabla_aB_{bc}+{\rm cyclic\ terms})$ 
is the gauge invariant quantity and determines the classical equation of 
motion. The static spherically symmetric solution   of an axion field 
interacting 
with gravity is the Schwartzschild metric supplemented by an axion 
potential  with a nonzero component given by \bow\ $B_{\theta\phi} 
=\alpha'Q\sin\theta/2$ where $Q$ is a constant. It 
follows that $H=0$ and no 
classical measurement outside the black hole could determine the axion 
charge of a black hole.  However as pointed out in \bow\ if  
  two strings originating at the same spacetime point are 
made to interfere at another spacetime point with  the volume 
enclosed by two world sheets enclosing the black hole,  
the quantum phase difference, 
between 
two strings at 
the end of their trajectory will be proportional to  $Q$.
  Thus $Q$ becomes a new observable quantity of the system 
despite the fact that strings world sheet do not touch the region of 
nonzero $H$.  This observable could be 
considered a new kind of quantum hair for black holes.

The  Anharonov-Bohm type 
experiment described above for strings would be difficult to perform.  If 
the axion field were coupled to another auxillary field, such as 
the dilaton, one might infer axion 
charge by measuring the auxillary field.   Interestingly one can give 
such an interpretation to  an  effective 
spacetime action describing the interaction between gravity, the axion and 
the dilaton derived by Giddings  et.al.\gidd . In their work  an exact 
classical solution of bosonic  
string theory is found as a conformally invariant sigma 
model, containing both the symmetric metric coupling, and  an
antisymmetric coupling $B_{\theta\phi}= \alpha'Q\sin\theta /2$. A fixed 
point of that theory occurs when $Q=\pi$.
These exact solutions found in \gidd \ can be interpreted as a black 
hole geometry without an asymptotically flat 
region.  
As such their solution cannot describe a black hole found in our spacetime.
However, on large scales the system can be studied using a spacetime
action with the world-sheet instanton effects included by an effective
interaction \gidd . In the rest of the paper we investigate the 
existence of asymptotically flat solutions using this spacetime action.

\newsec{Equations of Motion and Boundary Conditions}

\noindent 
We will write the static spherically symmetric Einstein metric as
\eqn\metric{ ds^2=-f(r)dt^2+{dr^2\over A(r)} +R^2 (r)d\Omega^2}
The four-dimensional low-energy effective action obtained in \gidd \ for 
$Q=\pi$ is $$
S=\int \sqrt{-g}\left[-{\cal R}+2(\nabla\phi)^2+V(\phi)\right]d^4x
$$
where $\phi$ is the dilaton field and $V$ reproduces the
effect of the the world sheet
instantons which couple to the topological gauge potential 
$B_{\theta\phi}$, 
\eqn\potential{
V(\phi)={2C\over r^2}e^{2\phi-2R^2  (r)/\alpha'}
}
Note that we are using the Einstein metric, whereas in \gidd \ the string
metric is used, dimensionally reduced to the $t$-$r$ plane.

In \potential,
$C$ is a positive constant which comes from the determinant of
fluctuations about the instantons. Hence $C$ naturally arises 
as a small
parameter, and this suggests using a perturbative approach to look for
solutions which are almost Schwarzchild black holes, with a small but
nonconstant dilaton. We first find these solutions, and then turn to the
exact case.

 After choosing the gauge 
$R=r$, 
 we obtain three independent equations of motion, with 
 $r=x r_h$ 
\eqnn\Eina
\eqnn\Einb
\eqnn\Einc
$$\eqalignno
{Ax^2\phi''&=(1+x\phi')Ce^{2\phi-x^2r_h^2}-(1+A)x\phi'&\Eina\cr
A'x&=1-A-Ax^2\phi'^2-Ce^{2\phi-x^2r_h^2}&\Einb\cr
(\ln f)'&=2x\phi'^2+(\ln A)'&\Einc\cr
}$$
where primes  denote differentiation with respect to $x$, and $2/\alpha'$ 
has been scaled away.
 For asymptotically flat spacetime,
we  require
\eqn\eLarge{A,f\to 1-2M/r\quad{\rm as}\quad r\to\infty.}
Requiring an event horizon at $r=r_h$ ($x=1$) implies that 
\eqn\eHorizon{A_h=f_h=0} 
The subscript $h$ will  stand for quantities at $x=1$.
For  a solution we require that $\phi$ is differentiable at $r_h$.  The 
energy density, $\rho$, is 
given by  $4\pi\rho=-4\pi T^t_t=(A\phi'^2+Ce^{2\phi-r^2}/r^2)/2$. 
Finiteness of $\rho$ at $r_h$ puts no further condition on $\phi$.
The total energy in the dilaton field outside the horizon is finite, 
 $4\pi\int_{r_h}^\infty \rho r^2
\sqrt{f/A}\thinspace
dr <\infty$,  if $\phi'$ vanishes asymptotically at least as
fast as $1/r^\delta$ with $\delta>1.5$. By solving 
\Eina \ asymptotically, we find that $\delta=2$.  Thus
\eqn\philarge{\phi\to\phi_\infty+{\cal O}(1/r)\quad{\rm as}\quad r\to\infty}
together with equation \eLarge \ constitute 
sufficient conditions for  aysmptotic flatness and finite total energy.   
To get a unique solution we must impose on a
particular value of $\phi$ at infinity.  We can  require that 
\eqn\ezero{\phi_\infty=0}
though we will see in section 4 that if $C$ is fixed there is a range 
of $\phi_h$ and hence a range of $\phi_\infty$ that leads to the 
asymptotically flat solutions. Alternatively we can impose \ezero, 
without loss of generality,  for all 
 solutions and let $C$ vary in which case there will be a range of values 
of $C$ that give \ezero.  That there is 
one-to-one correspondence between the two  parameters, 
$\phi_\infty$ and $C$, can be easily seen from 
the fact that setting $C=1$ amounts to redefining $\phi$ according to the 
scheme $Ce^{2\phi}\to e^{2\phi}$.  
Each  of the two parameters that one allows to vary 
has its own advantage. If one keeps $\phi_\infty$ fixed and 
lets $C$ vary, one can imagine turing on the perturbation potential $V$ 
off the Schwartzschild black hole.  But to discuss the existence of 
solution numerically it is more convenient to fix $C$ and let 
$\phi_\infty$ (or equivalently $\phi_h$) vary.

\newsec{Perturbative Solutions}

\noindent
We observe first that when $C=0$ the 
equations of motion for the metric functions are 
solved by the Schwartzschild spacetime.  
The equation for $\phi$ is that of a massless scalar field  minimally 
coupled to gravity and is solved by the unique solution $\phi=$ constant.  
For small $C$   we  look for a perturbative solution to the equations of 
motion.  Non-existence of a
perturbative 
solution would indicate that no exact solution exists while  existence 
 means only that the exact solution may exist.

Let the metric  be that of  Schwartzschild with $f=A=1-1/x$ and  $C$ and 
$\phi$ to be  small quantities of the same order. The equation of motion 
for $\phi$, $2\nabla^2\phi=V$, after 
 scaling $r=xr_h$, becomes
$$
[(x^2-x)\phi']'=Ce^{2\phi-x^2r_h^2}
$$
which can be immediately integrated to obtain 
$\phi$ to first order in the small quantity  
$$
\phi=C\int^x_\infty {G(t)\over t^2-t}dt
$$ 
where $G(z)=\int_1^z  e^{-r_h^2\zeta^2}d\zeta$. This solution 
satisfies all 
boundary conditions: ${d\phi\over dx}\sim 1/x^2$,
 $\phi\sim 0$ as 
$r\to\infty$, and  $\phi$  is bounded and 
differentiable at $r_h$. It is also consistent with our assumption that 
$\phi$ and $C$ are of the same order in the small quantity.  

One can also compute the first order correction to the metric. If one  
lets \break
$A=1-(1+C\delta)/x$ then one can solve for $\delta$  using \Einb.  
One finds $$\delta=G(x)$$
 Since $\lim_{x\to\infty}G(x)=$ positive constant,
this solution indicates that the presence of the dilaton 
field  increases  the ADM mass from the 
Schwartzschild value, a fact supported by the numerical solution found in
 the next section. From equation \Einc\ it follows that $f=A$ to first 
order in $C$.

\newsec{Numerical Results}

\noindent

For all cases considered in
this section, we take $r_h=C=1$ unless otherwise specified.  For all 
other values of $r_h$  we have tested,  
we  find that numerical solutions also exist, but we do not have
a proof that this is true in general.  At the end of this section we 
discuss the parameter space  of fixing $\phi_\infty$ and varying $C$.

Because equations \Eina \ and \Einb \ do 
not involve $f$ we solve these two equations for $A$ and  $\phi$ and 
then use \Einc \ to solve for $f$.  
Equations \Eina \ and \Einb \ are solved numerically with the boundary
conditions
\eLarge, and \eHorizon \ using the shooting method \press.  
 To start shooting, initial values must be chosen for
$A$, $\phi$, and $\phi'$ at $r_h$.  But
 at $x=1$,
equation
\Eina \ gives a constraint between $\phi_h$ and
$\phi'_h$:
\eqn\constraint{\phi'_h=1/( e^{r_h^2-2\phi_h}-1)   }
 Note that the solution in section 2 satisfies this constraint.
For a given value of $r_h$ we thus have only one  shooting parameter,
$\phi_h$.   Equation \constraint\ is singular when $2\phi_h=r_h^2$.  
Indeed, a   solution exists only when $2\phi_h<r_h^2$,
because the equation \Einb\ gives 
\eqn\Aprime{A'_h= 1 - e^{2\phi_h-r_h^2}}
and since $A_h=0$ it is necessary that $A'_h>0$ which is possible only if 
$2\phi_h>r_h^2$. 

One sees that  $\phi$ monotically 
increases with $x$ as follows.
For $2\phi_h>r_h^2$,  $\phi'_h>0$. Thus if there exists a local extremum 
of $\phi$ for some $x=x_e>1$ it must be a local maximum.  However 
according to equation \Eina\ 
$$
A(x_e)x_e^2\phi''(x_e) =e^{2\phi(x_e)-x_e^2r_h^2}>0
$$
i.e.~that local extremum can only be a local minimum.  Therefore there 
can be no local extremum of $\phi$ for any finite value of $x$.

\centerline{\epsfbox{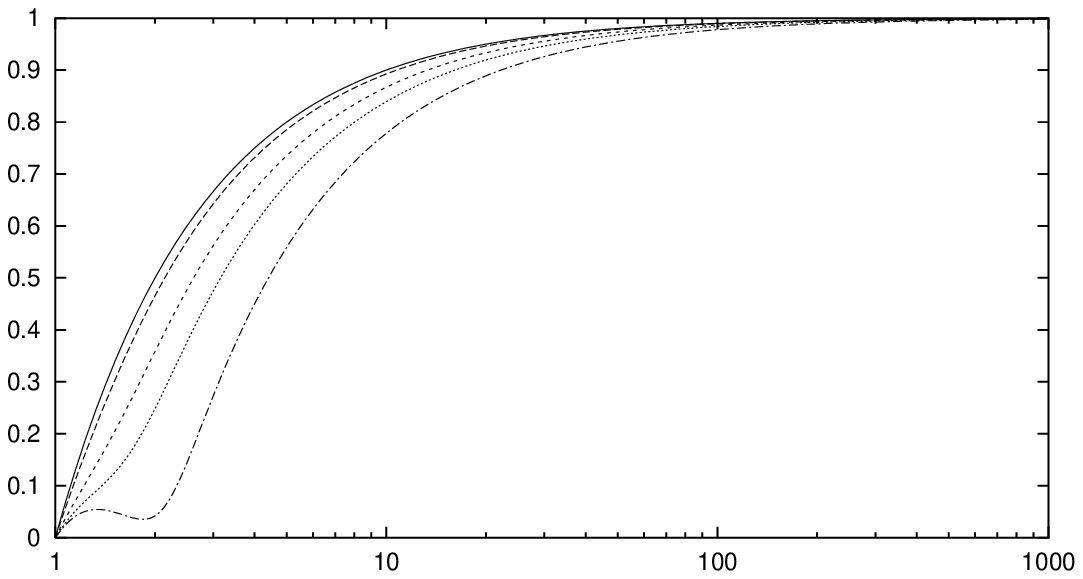}}
\centerline{\it Figure 1.\quad $A$ vs.~$x$ \quad\quad $(r_h,C)=(1,1)$}
\midinsert\narrower\narrower\narrower\noindent
{\sevenrm  \baselineskip=5pt plus 1pt minus 1pt
Solid line is the 
Schwartzschild solution, $1-1/x$.  For the dashed 
lines, from left to right, 
$\phi_h= -0.4, 0.1, 0.2, 0.235$.} 
\endinsert\vskip3mm 

We find that for 
each $r_h$
there is a range of $\phi_h\in(-\infty,\phi_h^{\rm max}]$ that will give 
$\phi_\infty=$constant.    For example for $r_h=1$, 
$\phi_h=-0.21678$ gives $\phi_\infty=0$ and $\phi_h^{\rm max}=0.235596$.  
The relation between $\phi_h^{\rm max}$ and $r_h$ is summarized in 
figure 7.  For 
all $\phi_h<\phi_h^{\rm max}$ $\phi$ monotically increases to a constant 
value and $A$ approaches the Schwartzschild result, $1-1/x$, 
asymptotically.    
Figure 1 also shows that $A$ approaches Schwartzschild solution 
everywhere as 
$\phi_h\to-\infty$.  This is expected because $\phi_h\to-\infty$ is the 
same as $C\to0^+$.

Once $A$ and $\phi$ are known $f$ can be  
integrated using equation \Einc. Note that \Einc\ is linear in 
$f$.  This is compatible with the fact that the time coordinate can be 
scaled 
by a constant. Because of the event horizon,  \Einc \ does not 
determine 
$f'_h$.  Initially we take $f'_h=A'_h$ which gives a solution 
such 
that $\lim_{r\to\infty}f=f_\infty$.  Further scaling of time coordinate 
is necessary to obtain $\lim_{r\to\infty}f=1$ \ i.e.~for the final 
solution we 
pick $f'_h=A'_h/f_\infty$. $f'_h$ is necessary for computing Hawking 
temperature later.

\vskip4mm
\centerline{\epsfbox{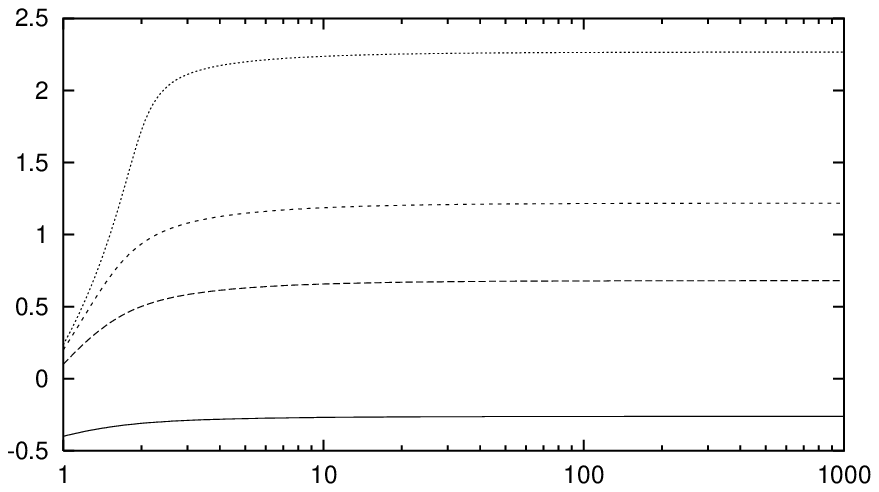}}
\centerline{\it Figure 2.\quad $\phi$ vs.~$x$ \qquad $r_h=1$}
\vskip4mm
For the numerical analysis it is  convenient to think of the solutions as 
being 
parametrized by $(r_h,\phi_h)$.  However, we may also consider the 
solutions as being parametrized by two global charges: the ADM mass, $M$, 
and the dilaton charge, $D$.  We now turn to calculating these two charges.

The ADM mass can be read off of the large $r$ behavior of $A$, $1-2M/r$.
Figure 3 shows that for a fixed $\phi_\infty$, $(M/M_{C=0})$ increases 
with $C$ as noted with the perturbative solution.

\vskip3mm
\centerline{\epsfysize=50mm\epsfbox{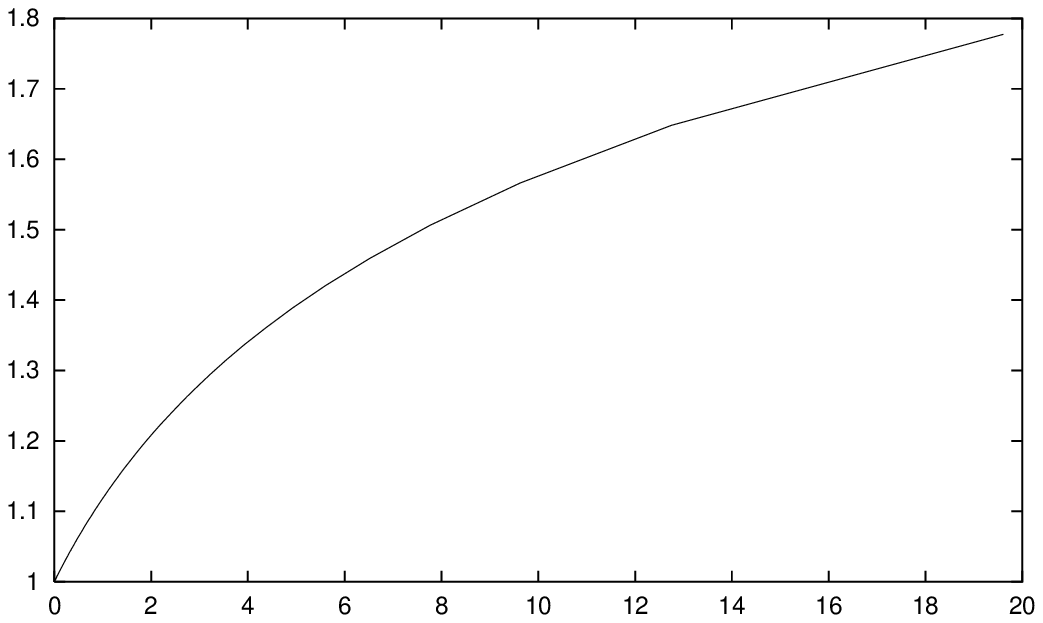}}
\centerline{\it Figure 3.\quad $(M/M_{C=0})$ vs.~$C$\qquad 
$(r_h,\phi_\infty)=(1,0)$ } \vskip4mm

A plot of 
$(x^2\phi')$ vs.~$x$ in figure 4 shows that $\phi'\sim 
{\rm constant}/x^2$ as $x\to\infty$. This asymptotic behavior agrees with 
that of perturbative solution and satisfies equation \philarge.
 With this behavior of $\phi'$,
 $$D={1\over 4\pi}\int d^2\Sigma^\mu \nabla_\mu\phi={1\over 4\pi}\int 
d\Omega\thinspace r^2 {d\phi\over dr}$$ 
is finite. 
  $D$ should not be thought of as  another free 
parameter. 
  Figure 5 shows one-to-one correspondence between $D$ and 
$\phi_\infty$.  Also there is one-to-one correspondence 
between two parameter spaces $(r_h,\phi_\infty)$ and $(M,D)$.

\vskip4mm
\centerline{\epsfysize=48mm\epsfbox{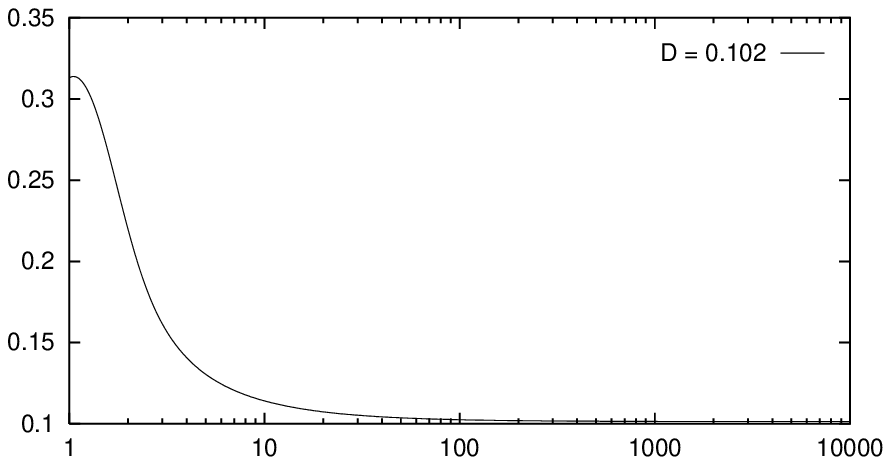}}
\centerline{\it Figure 4.\quad $\left(x^2\phi'\right)$ vs.~$x$ \qquad 
$(r_h,C,\phi_\infty)=(1,1,0)$}
\vskip4mm
\centerline{\epsfysize=48mm\epsfbox{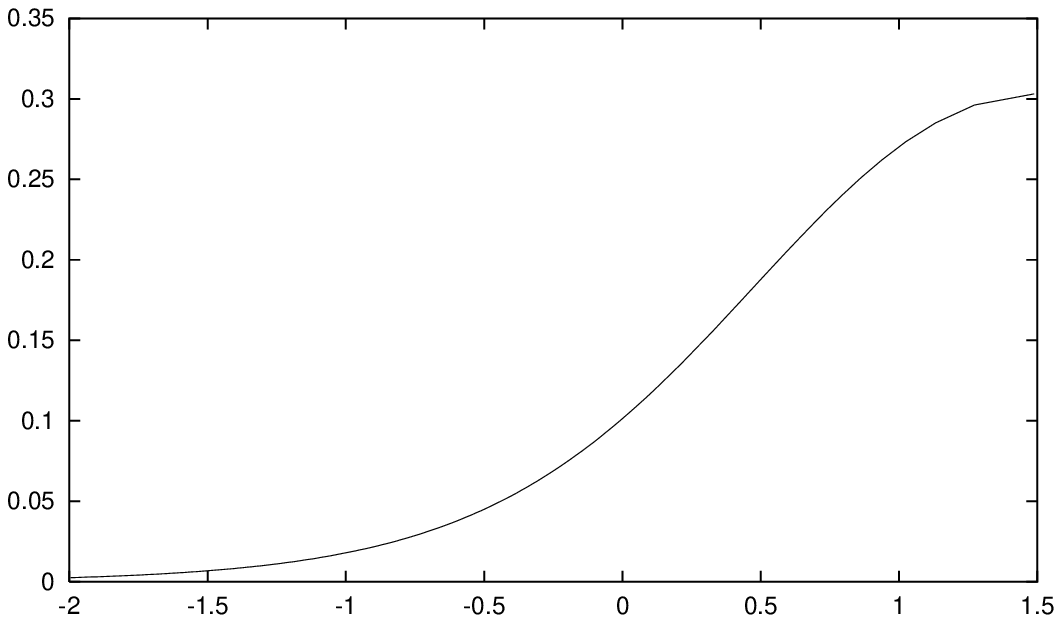}}
\centerline{\it Figure 5.\quad $D$ vs.~$\phi_\infty$\qquad $(r_h,C)=(1,1)$}
\vskip4mm

  Another quantity of interest is 
Hawking temperature, $T_H=\kappa/2\pi$. 
$\kappa$ is surface gravity \wald\ and is given by
$\lim_{r\to r_h}(Va)$ where $a=\sqrt{a_ca^c}$, 
$a^c=(\xi^b\nabla_b\xi^c)/(-\xi^d\xi_d)$, 
$\xi^c=\left({\partial\over\partial t}\right)^c$, and 
$V=\sqrt{-\xi^c\xi_c}$.  
We find that 
$$4\pi T_H=\lim_{r\to 
r_H}f'/\sqrt{f/A}=
\sqrt{f_h'A_h'}=A_h'/\sqrt{f_\infty}$$
 continuously decreases with $\phi_\infty$, and hence 
with $\phi_h$.  An alternative expression, obtained by using equation 
\Aprime,  $4\pi T_H= 
(1-e^{2\phi_h-r_h^2})/\sqrt{f_\infty}$
explains 
the decreasing of $T_H$ with $\phi_h$.  It also supports 
our finding that no solution exists for large value of 
$\phi_h$ because if 
$\phi_h>\phi_c\equiv r_h^2/2$, $T_H$ will be negative.
 Figure 7 shows the relation between 
 $\phi_h^{\rm max}$ and $r_h$ and it shows that as  $r_h\to\infty$, 
$\phi_h^{\rm max}\to r_h^2/2=$ Hawking temperture bound.

For a fixed $r_h$ the effect of dilaton charge is to lower the Hawking 
temperature as evident by figure 6.   In this respect the 
dilaton charge 
has a similar effect on the black hole as the electric or magnetic 
charge. 
Recall that the Hawking 
temperature of a Reissner-Nordstrom black hole is given by
$$
4\pi T_H = {r_+^2-Q_{\rm em}^2\over  r_+^3}
$$
where $r_+$ is the outer horizon radius, and $Q_{\rm em}$ the electric or 
magnetic charge of the black hole.

\vskip4mm
\centerline{\epsfysize=45mm\epsfbox{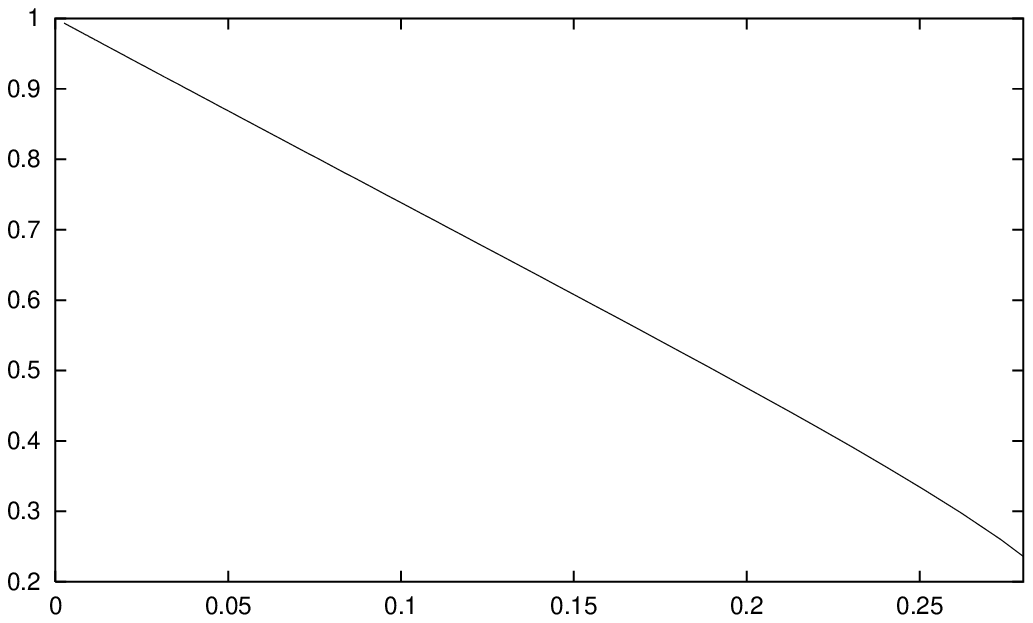}}
\centerline{\it Figure 6.\quad $4\pi T_H$ vs.~$D$\quad\quad $r_h=1$}

Throughout this section we have taken $C=1$ and found that the  solution 
exists when $\phi_h\in(-\infty,\phi_h^{\rm max}]$ (or 
equivalently $\phi_\infty\in(-\infty,\phi_\infty^{\rm max}]$.)  As noted 
at the end of section 
2 if $\phi$ is the solution with $C=1$ then $\tilde\phi$ will be 
 the solution 
for a different positive $C$ where $\tilde\phi=\phi-\ln\sqrt{C}$.  If we 
demand that $\tilde\phi_\infty=0$ always, then for a fixed $r_h$ the 
solution exists if and only if $C\in[0,C^{\rm max}]$ where $C^{\rm 
max}=e^{2\phi_\infty^{\rm max}}$.  It can be easily verified that 
 picking $\tilde\phi_h=\phi_h-\phi_\infty$ and 
$C=e^{2\phi_\infty}$ will ensure $\tilde\phi_\infty=0$.

\newsec{Consistency with No-Hair Theorems}

\noindent
To check whether black hole solutions we found are consistent with 
various no-hair theorems we review the proofs such theorems.
Consider a theory of scalar field interacting 
with gravity described by the action,
$S=\int [{\cal -R}+(\nabla\phi)^2/2+V(\phi,x)]\sqrt{-g}d^4x$.
If we write the line element  in the form of \metric\ with $R=r$
 the equation of motion  is $
\sqrt{A/ f}\left(r^2\phi'\sqrt{fA}\right)'
=r^2\partial V/ \partial \phi$
where primes denote $d/dr$. Now for asymptotic flatness we can 
 consistently require 
 $\lim_{r\to\infty}\phi=0$.  Multiplying 
both sides by $\phi$ and integrating by parts gives 
$\phi\phi'r^2\sqrt{fA}\big|_{r_h}^\infty=\int_{r_h}^\infty 
r^2\sqrt {fA}\left[\phi'^2+{\partial V\over 
\partial \phi}\phi/A\right]dr$.  The left side is zero 
on account of the boundary conditions imposed at $r_h$ and $\infty$.
 If ${\partial V\over \partial 
\phi}\phi$ 
is positive definite for $r>r_h$, then the integrand is 
positive definite and we see that for such forms of $V(\phi)$ the only 
solution 
for $\phi$ is the trivial one: $\phi\equiv 0$.  However, for the interaction 
$V$ we 
use $\phi{\partial V\over \partial\phi}$ is not necessarily positive 
definite.

No-hair theorems using scaling arguments \heusler\  impose slightly 
different condition, namely, that $V$ itself be positive definite. It 
appears at first sight that $V$ we use satisfies this condition.  However
the proof  in \heusler\ makes use of the fact that 
$V$ does not depend on coordinates explicitly which our $V$ clearly does not 
satisfy.

\newsec{Stability Analysis}

\noindent
To study the stability of the solution,
we will  consider spherically symmetric 
perturbations 
around the static solution. We willl show that  the solution is stable 
for sufficiently small $C$. Let
$$
\eqalign{
 f(r,t)&=f_0(r) + \delta f(r,t)\cr
 A(r,t)&=A_0(r) +  \delta A(r,t)\cr
 \phi(r,t)&=\phi_0(r) + \delta \phi(r,t)\cr
Q(r,t)&=\pi +  \delta Q(r,t).\cr}
$$ 
$f_0$, $A_0$, and $\phi_0$ are the static solution found in the previous 
sections and $\delta$ denotes  smallness compared to static solution.  
Here we are making a perturbation around the topological charge
$Q$ as well, which in the static case was taken to be $\pi$.  
The new action is
$$
S=\int \left[-{\cal R} + 2(\nabla \phi)^2 + e^{-4\phi}{H^2\over 
3} - {2C\over r^2}e^{2\phi-2r^2/\alpha'}\cos Q 
\right]\sqrt{-g}d^4x $$
where  $H_{abc}$ has nonzero components
$$
H_{r\theta\phi}=\alpha'{\sin\theta\over 2}\partial_r\delta Q
\quad\quad{\rm and}\quad\quad
H_{t\theta\phi}=\alpha'{\sin\theta\over 2}\partial_t\delta Q
$$

The linearized gravitational field equations obtained from the action do not 
lead to an
unstable mode because a spherically symmetric gravitational field has no 
dynamical degrees of freedom.   Explicit demonstration of this fact is 
given in \sz.

After dropping the subscript  $0$ of the background solution, 
 the linearized equation of motion for 
$\delta\phi$ is, 
$$
\sqrt{A\over f}\left(x^2\sqrt{fA}\delta\phi'\right)'-{x^2\omega^2\over 
f}\delta\phi={2 C\over x^2}\delta\phi e^{2\phi-x^2r_h^2}
 $$
where $'=\partial/\partial x$  $(x=r/r_h)$ and $\alpha'$ has been scaled
away.
We have also scaled the time coordinate by a constant and taken the time 
 dependence of $\delta\phi$ to be $ e^{\omega t}$.  
Then, after transforming to the tortoise coordinate ${dy\over 
dx}={1\over\sqrt{fA}}$, we obtain the equation for $\psi(y)\equiv 
x\delta\phi$ as 
$$
{d^2\psi\over dy^2} + (E-V_\phi)\psi=0
$$
where $E=-\omega^2$, and 
$$
V_\phi= {1\over 2x}{d\over dx}(fA) + {2fC\over x^4}e^{2\phi-x^2r_h^2}
$$
 This is one dimensional nonrelativistic schrodinger equation for a 
particle 
with mass $={1\over 2}$ moving under the influence of the potential 
$V_\phi$.  
Instability of the solution corresponds to the existence of the bound 
states,  $E<0$, solutions of the schrodinger equation.  But  
 $V_\phi$ is positive definite for most of the parameter space of the 
static
solution. See figure 7. Thus in this region of parameter space, where $C$ 
and/or $r_h$ are small, a linear time dependent perturbation in the 
dilaton field leads to no instability.\foot{We believe that even in
the region of parameter space which makes $V_\phi$  negative 
no bound state with $E<0$ is  possible.  The reason is that  when $V_\phi$ 
does become 
negative it is in the form of a shallow well surrounded by a  high 
barrier and a small barrier.}

\vskip3mm
\centerline{\epsfysize=50mm\epsfbox{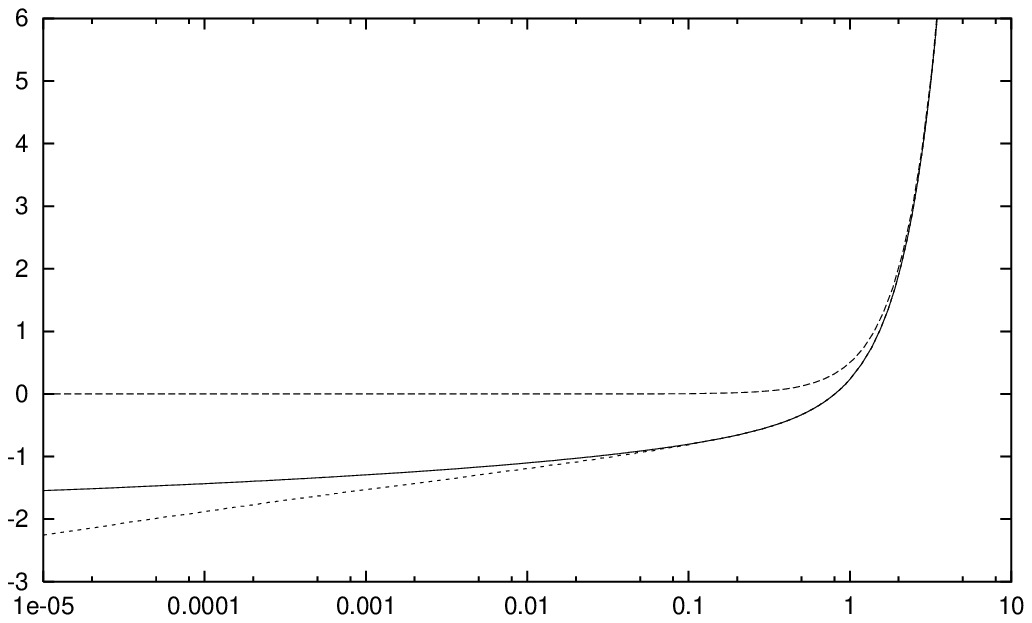}}
\centerline{\it Figure 7.\quad $\phi_h$ vs.~$r_h$ }
\midinsert\narrower\narrower\narrower
\noindent  {\sevenrm 
 \baselineskip=5pt plus 1pt minus 1pt
According to the remarks at the end of section 5 
this figure can be looked upon as $C$ vs.~$r_h$ with $\phi_\infty=0$.  
Dashed line is 
the curve $r_h^2/2$, Hawking temperature bound on $\phi_h$ with $C=1$.
Solid line is $\phi_h^{\rm max}$ vs.~$r_h$; it determines the boundary of 
solution space i.e. no 
static solution exists for  any point above the solid line.  $V_\phi$ is 
positive definite below the small-dashed line.}
 \endinsert

We next
investigate the equation of motion for $\delta Q$ in the parameter space 
in which $V_\phi$ 
is positive definite.  Assuming the same time dependence as before the 
linearized equation of motion for $\delta Q$ can also be put into 
the form of the schrodinger equation as before with the  potential energy 
function
$$
V_Q(y) = Af\left[B^2-B'(x)\right]-{B\over 2}
{d\over dx}(fA)-3fx^2r_h^4Ce^{6\phi-x^2r_h^2}
$$
where $B(x) = 2\phi'(x)+1/x$.  Here we have set $\alpha'=2$ for 
convenience. 
 To study $V_Q$ it is convenient to think\foot{
Because $Ce^{6\phi}$ appears in $V_Q$ there is no simple 
 tranformation between the parameter space  where $C$ is 
fixed and $\phi_\infty$ varied and that where $\phi_\infty$ is fixed and 
$C$ varied.  However, when one considers which part of the  
{\it static} 
solution space is stable under perturbation, it is  consistent to 
use the simple transformation rule given at the end of section 5.      }
of the representation in 
which $(r_h,C)$ is varied and $\phi_\infty=0$.  One should  imagine 
figure 7 as $C$ vs.~$r_h$ with  $\phi_h\to-\infty$ being $C\to0$.  Now the 
last term in $V_Q$ will be small  when $C$ and/or $r_h$ 
are small whereas other terms do not depend explicitly on $C$ or $r_h$. 
Thus the effect of the last term in $V_Q$ will show up for the points in 
upper right of figure 7.  Figure 8 shows a 
typical $V_Q$ when $C$ and/or $r_h$ is small; it has a well --- which we 
will refer to as the ``primary'' well --- followed by a barrier.
From now on we will  consider varying $(r_h,\phi_h)$.

\centerline{\epsfysize=35mm\epsfbox{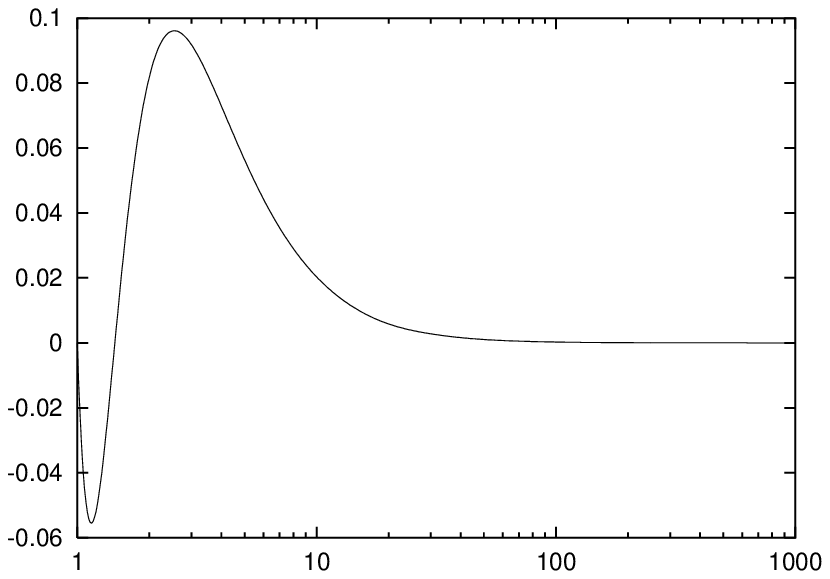}}
\centerline{\it Figure 8.\quad $V_Q$ vs.~$x$\quad\quad 
$(r_h,C,\phi_h)=(0.077161,1,-1)$}

\noindent
    The 
effect of  $r_h$ and $\phi_h$ when they are larger (i.e.~when they are in 
the upper right of figure 7)  is 
to introduce a ``secondary" well whose location depends on $r_h$.   In 
general 
this secondary well gets deeper as $r_h$ and $\phi_h$ gets bigger and moves 
to the left as $r_h$ gets bigger.  
 See figures 11abcd. If there is any bound state of 
$V_Q$ the particle will be localized in the primary well or the 
secondary well.  Since big $r_h$ and $\phi_h$ give a deep secondary 
well, we start  looking for bound states for such cases.

 To find the bound states of  $V_Q$ 
it is numerically simpler to use the original $x$ coordinate.
 As $x\to\infty$ the solution for $\delta Q$  is 
$e^{-\omega x}$ and 
 the linearized
equation of $q\equiv\delta Q e^{\omega x}$ is 
$$
q''-(a+2\omega)q'+(\omega^2+\omega a-b)q=0
$$ 
where $'=d/dx$, 
$$
a= 2B(x)-{1\over 2}\ln'(fA)\quad\quad\hbox{and}\quad\quad
b={\omega^2\over fA}-{3x^2r_h^4\over A}Ce^{6\phi-x^2r_h^2}
$$
The wave function in the schrodinger equation is related to $q$ via
$\psi=(q/ x)e^{-2\phi-\omega x}$  .

For arbitrary value of $\omega$ and arbitrary potential 
$V_Q$, no regular solution of $\psi$ will be possible. We call $\psi$ a 
regular  solution if it goes to zero sufficiently fast as $x\to 1$ and $x\to 
\infty$.  We 
find that for a deep enough secondary well of $V_Q$ and a discrete set of 
positive values of $\omega$ regular solutions exist.  These regular 
solutions are the bound states we are looking for. 

\centerline{\epsfysize=48mm\epsfbox{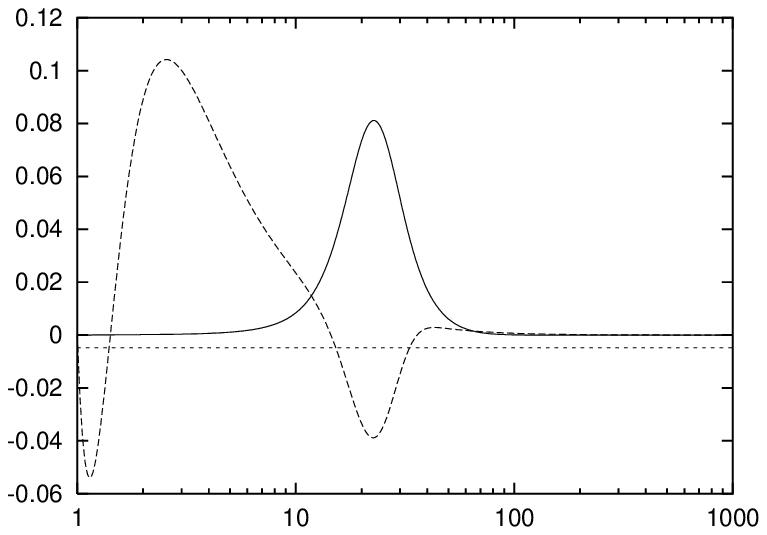}}
\centerline{\it Figure 9.\quad $\psi$ and $V_Q$ vs.~$x$ \quad\quad 
$(r_h,C,\phi_h)=(0.077161,1,-0.875)$}
\centerline{\sevenrm Solid line is the wave function of the ground 
state and  horizontal line  the energy level.}

\noindent
As shown in figure 9 the wave function is localized around the secondary 
well, and this is 
evidence that no bound state will be possible when the secondary well is 
shallow enough.  Therefore for a fixed $r_h$ there may be bound states 
when $\phi_h$ is close to $\phi_h^{\rm max}$ but as $\phi_h^{\rm max}\to 
-\infty$ bound states will 
disappear.  For a fixed $r_h$ we can graph $\omega$ of the ground state 
against $\phi_h$ and by extrapolating find the zero of that graph, 
$\phi_h^0$. See figure 10.
  For $\phi_h<\phi_h^0$ no bound state is 
possible and hence the solution is stable.

\newsec{Summary}

\noindent

We have presented the evidence for the existence of scalar hair of a 
black hole in the presence of string instantons which couple to a 
topological gauge potential.  The space time action that we use includes 
an effective interaction due to the string instantons wrapping around the 
Euclidean black hole.  Measurement of the dilaton charge of the black 
hole can be looked upon as an indirect detection of the topological axion 
charge of the black hole.  For example, if a ``dilatonic ball'' held at a 
large distance from the black hole experiences an attractive force 
towards the black hole in excess of the gravitational force the dilatonic 
charge of the black hole will be ascertained, and hence the axionic charge.
\vskip3mm
%\vfill\eject%\blaa
\noindent
{\bf Acknowledgements}

\noindent

JT would like to thank D.~Kastor and S.~Giddings for useful 
conversations.  This work is supported in part by  NSF Grant 
NSF-THY-8714-684-A01.

\vskip7mm
\centerline{\epsfysize=55mm\epsfbox{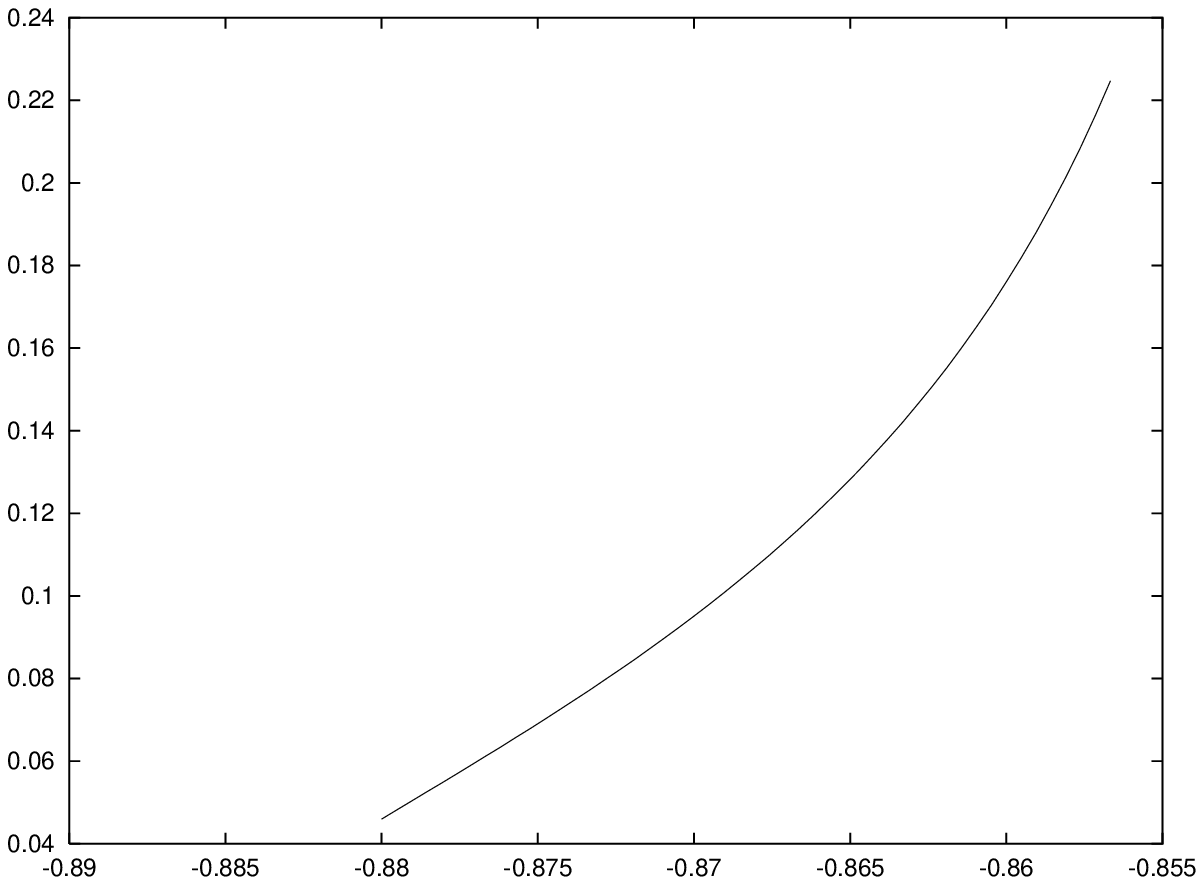}}
\centerline{\it Figure 10.\quad $\omega$ vs.~$\phi_h$  \quad 
$(r_h,C)=(0.077161,1)$  $\phi_h^0\approx -0.892$ } \midinsert\narrower\narrower\narrower
\noindent{\sevenrm
 \baselineskip=5pt plus 1pt minus 1pt
 Because there is one-to-one  correspondence 
between $C$ and $\phi_h$ (of the static solution) this figure can be 
roughly  looked upon as $\omega$ 
vs.~$C$ with $\phi_\infty=0$. }
\endinsert
\vfil\eject

\centerline{\epsfysize=40mm\epsfbox{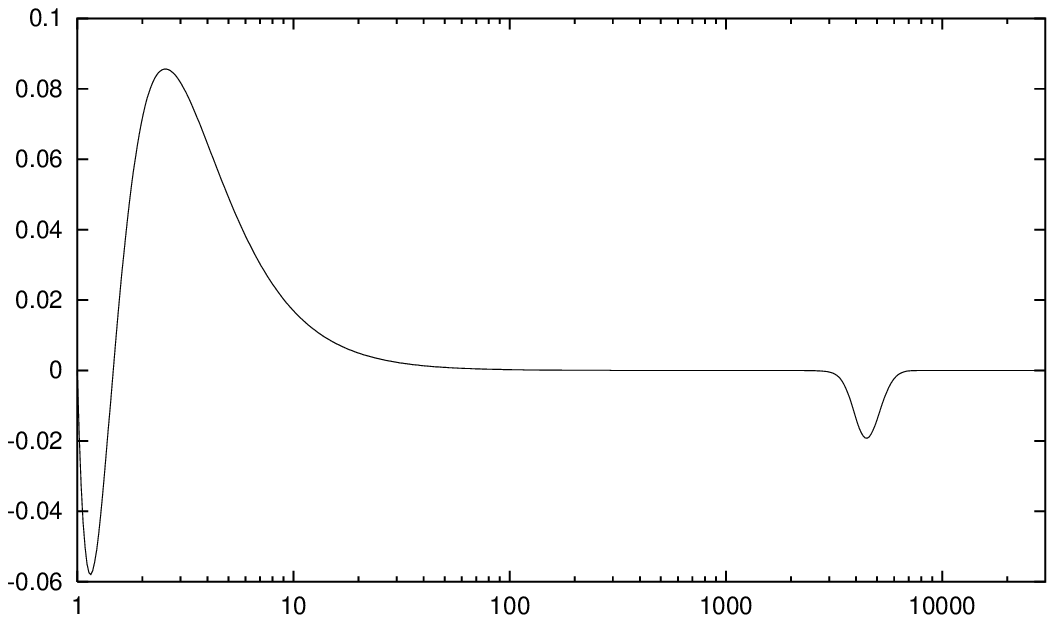}}
\centerline{\it Figure 11a. \quad $V_Q$ vs.~$x$\quad\quad 
$(r_h,C,\phi_h)=(0.0005265,1,-1.33594)$}

\centerline{\epsfysize=40mm\epsfbox{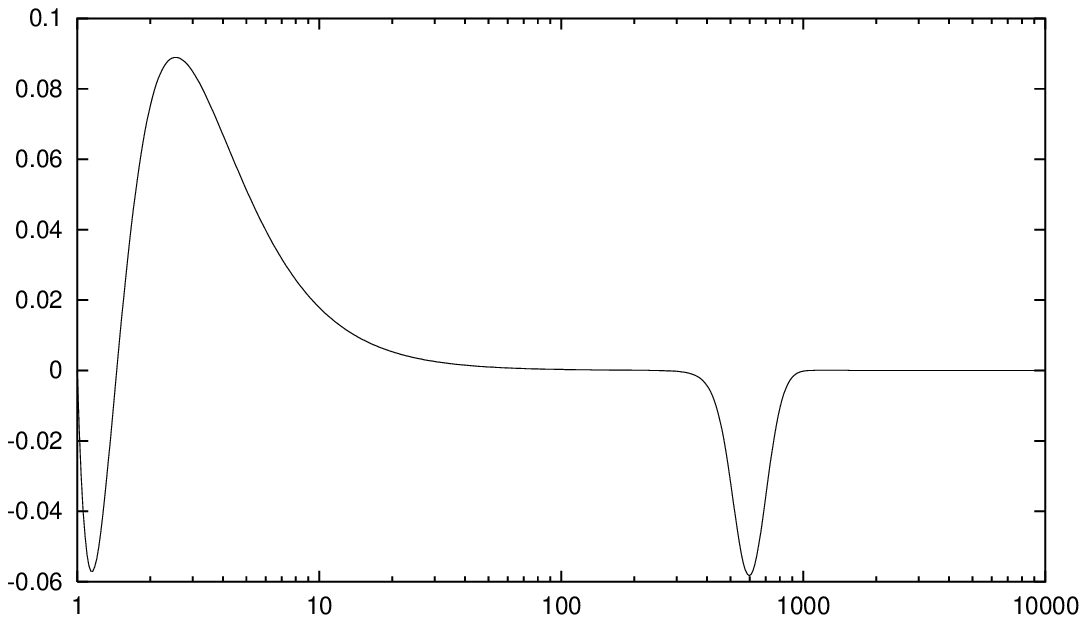}}
\centerline{\it Figure 11b. \quad $V_Q$ vs.~$x$\quad\quad 
$(r_h,C,\phi_h)=(0.003679,1,-1.194)$}

\centerline{\epsfysize=40mm\epsfbox{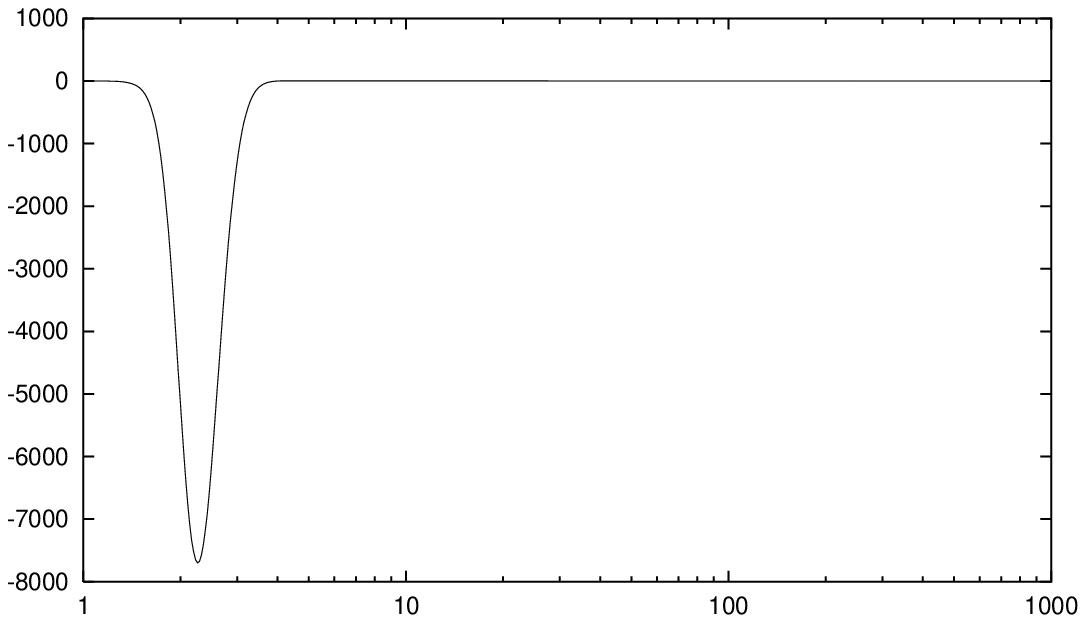}}
\centerline{\it Figure 11c. \quad $V_Q$ vs.~$x$\quad\quad 
$(r_h,C,\phi_h)=(1,1,0.23)$}
\centerline{$\phi_h$'s of figures a,b,c lie on the small dashed line of 
figure 7.}

\centerline{\epsfysize=40mm\epsfbox{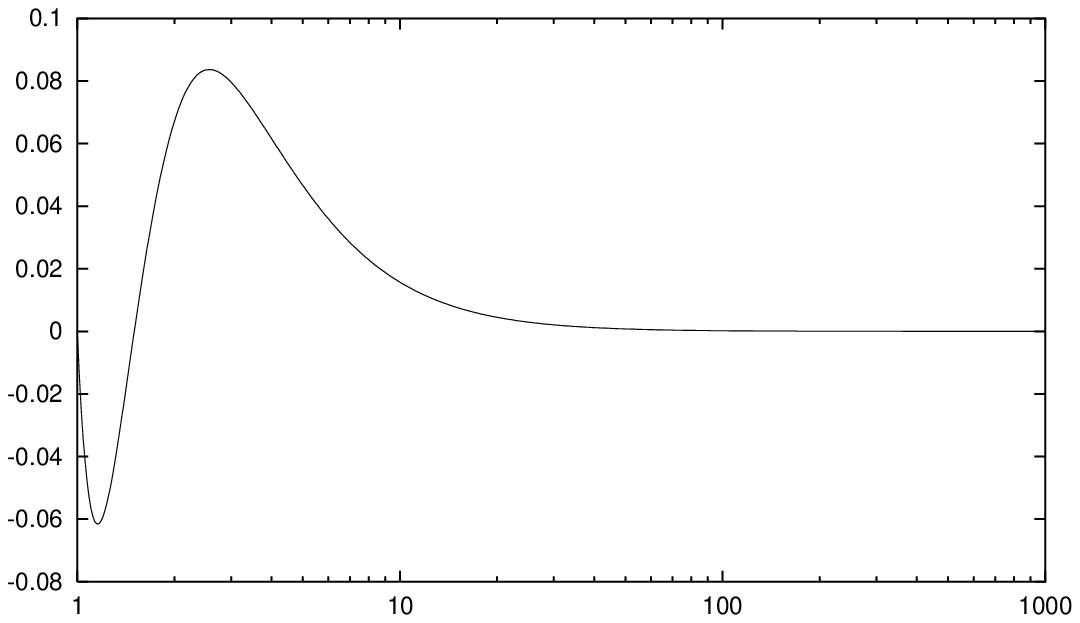}}
\centerline{\it Figure 11d. \quad $V_Q$ vs.~$x$\quad\quad 
$(r_h,C,\phi_h)=(1,1,-1.194)$}
\vfil\eject

\listrefs

\bye